\documentclass[10pt]{emulateapj}

\shorttitle{Solar Wind Magnetic Helicity}
\shortauthors{Howes and Quataert}

\newcommand\Alfven{Alfv\'en }
\newcommand{\V}[1]{\mathbf{#1}} 
 
\newcommand{\zhat}{\mbox{$\hat{\mathbf{z}}$}} 
\newcommand{\xhat}{\mbox{$\hat{\mathbf{x}}$}} 
\newcommand{\yhat}{\mbox{$\hat{\mathbf{y}}$}} 
\newcommand{\figref}[1]{Figure~\ref{#1}}

\newcommand{\eqref}[1]{equation~(\ref{#1})}
\newcommand{\eqsref}[2]{equations~(\ref{#1})~and~(\ref{#2})}

\begin{document}


\title{On the Interpretation of Magnetic Helicity Signatures in the
  Dissipation Range of Solar Wind Turbulence}


\author{Gregory ~G. Howes}
\affil{Department of Physics and Astronomy, University of
Iowa, Iowa City, IA, 52242}
\and
\author{Eliot~Quataert}
\affil{Department of Astronomy, University ofCalifornia, Berkeley, CA, 94720}
 


\begin{abstract}
  Measurements of small-scale turbulent fluctuations in the solar wind
  find a non-zero right-handed magnetic helicity. This has been
  interpreted as evidence for ion cyclotron damping.  However,
  theoretical and empirical evidence suggests that the majority of the
  energy in solar wind turbulence resides in low frequency anisotropic
  kinetic \Alfven wave fluctuations that are not subject to ion
  cyclotron damping.  We demonstrate that a dissipation range
  comprised of kinetic \Alfven waves also produces a net right-handed
  fluctuating magnetic helicity signature consistent with
  observations. Thus, the observed magnetic helicity signature does
  not necessarily imply that ion cyclotron damping is energetically
  important in the solar wind.
\end{abstract}



\section{Introduction}
The identification of the physical mechanisms responsible for the
dissipation of turbulence in the solar wind, and for the resulting
heating of the solar wind plasma, remains an important and unsolved
problem of heliospheric physics. An important clue to this problem is
the observed non-zero fluctuating magnetic helicity signature at
scales corresponding to the dissipation range of solar wind
turbulence.

\citet{Matthaeus:1982a} first proposed the ``fluctuating'' magnetic
helicity as a diagnostic of solar wind turbulence, defining the
``reduced fluctuating'' magnetic helicity spectrum derivable from
observational data (see \S \ref{sec:red} below).  A subsequent study,
corresponding to scales within the inertial range, found values that
fluctuated randomly in sign, and suggested an interpretation that ``a
substantial degree of helicity or circular polarization exists
throughout the wavenumber spectrum, but the sense of polarization or
handedness alternates randomly'' \citep{Matthaeus:1982b}. Based on a
study of the fluctuating magnetic helicity of solutions to the linear
Vlasov-Maxwell dispersion relation, \citet{Gary:1986} suggested
instead that, at inertial range scales, all eigenmodes have a very
small {\it intrinsic} normalized fluctuating magnetic helicity,
eliminating the need to invoke an ensemble of waves with both left-
and right-handed helicity to explain the observations.

Subsequent higher time resolution measurements, corresponding to
scales in the dissipation range, exhibited a non-zero net reduced
fluctuating magnetic helicity signature, with the sign apparently
correlated with the direction of the magnetic sector
\citep{Goldstein:1994}. Assuming dominantly anti-sunward propagating
waves, the study concluded that these fluctuations had right-handed
helicity. The proposed interpretation was that left-hand polarized
Alfv\'en/ion cyclotron waves were preferentially damped by cyclotron
resonance with the ions, leaving undamped right-hand polarized
fast/whistler waves as the dominant wave mode in the dissipation
range, producing the measured net reduced fluctuating magnetic
helicity. We refer to this as the \emph{cyclotron damping
  interpretation}.

A subsequent analysis of more solar wind intervals confirmed these
findings for the dissipation range \citep{Leamon:1998a}.
\citet{Leamon:1998b} argued that a comparison of the normalized
cross-helicity in the inertial range (as a proxy for the dominant wave
propagation direction in the dissipation range) to the measured
normalized reduced fluctuating magnetic helicity provides evidence for
the importance of ion cyclotron damping, which would selectively
remove the left-hand polarized waves from the turbulence; using a
simple rate balance calculation, they concluded that the ratio of
damping by cyclotron resonant to non-cyclotron resonant dissipation
mechanisms was of order unity. A recent study performing the same
analysis on a much larger data set concurred with this conclusion
\citep{Hamilton:2008}.

In this Letter, we demonstrate that a dissipation range comprised of
kinetic \Alfven waves produces a reduced fluctuating magnetic helicity
signature consistent with observations.  A dissipation range of this
nature results from an anisotropic cascade to high perpendicular
wavenumber with $k_\perp \gg k_\parallel$; such a cascade is
consistent with existing theories for low-frequency plasma turbulence
\citep{Goldreich:1995,Boldyrev:2006,Howes:2008a,Schekochihin:2009}, numerical
simulations \citep{Cho:2000,Howes:2008b}, and observations in the
solar wind \citep{Horbury:2008,Podesta:2009a}.  Our results imply
that no conclusions can be drawn about the importance of ion cyclotron
damping in the solar wind based on the observed magnetic helicity
signature alone.

\section{Fluctuating Magnetic Helicity}
\label{sec:hm}
The magnetic helicity is defined as the integral over the plasma
volume $H_m \equiv \int d^3\V{r} \V{A} \cdot \V{B}$, where $\V{A}$ is
the vector potential which defines the magnetic field via $\V{B}
=\nabla \times \V{A}$. This integral is an invariant of ideal
Magnetohydrodynamics (MHD) in the absence of a mean magnetic field
\citep{Woltjer:1958a,Woltjer:1958b}. \citet{Matthaeus:1982b} chose to
set aside the complications associated with the presence of a mean
magnetic field, defining the \emph{fluctuating magnetic helicity} by
$H'_m= \int d\V{r} \delta \V{A} \cdot \delta \V{B}$, where the
fluctuating quantities denoted by $\delta$ do not include
contributions from the mean field.

Modeling the turbulent magnetic field\footnote{We assume that
turbulent fluctuations are reasonably modeled as a collection of
linear wave modes. Nonlinear interactions, neglected here, will serve
to replenish energy lost from wave modes, so we neglect the linear
wave damping and take only the real frequency.} by
\begin{equation}
\V{B}(\V{r},t)= B_0 \zhat +
\sum_{\V{k}}\V{B}(\V{k})e^{i(\V{k}\cdot \V{r} - \omega t)}
\label{eq:b}
\end{equation}
in a periodic cube of plasma with volume $L^3$, we obtain $H'_m= L^3
\sum_{\V{k}}H'_m(\V{k})$, where the
\emph{fluctuating magnetic helicity density} for each wave vector
$\V{k}$ is defined by $H'_m(\V{k}) \equiv \V{A}(\V{k}) \cdot
\V{B}^*(\V{k})$. Here $\V{B}(- \V{k})
= \V{B}^*(\V{k})$ and $\omega(-\V{k}) = -\omega^*(\V{k})$ are reality
conditions and $\V{B}^*(\V{k})$ is the complex conjugate of the
Fourier coefficient.  Specifying the Coulomb gauge $\nabla \cdot
\V{A}=0$, we obtain
\begin{eqnarray}
H'_m(\V{k})
&=& i \frac{B_yB^*_z-B^*_yB_z}{k_x}
= i \frac{B_zB^*_x-B^*_zB_x}{k_y} \nonumber \\
&=&i \frac{B_xB^*_y-B^*_xB_y}{k_z} 
\label{eq:hm},
\end{eqnarray}
where the components $B_j(\V{k})$ arise from the eigenfunctions of the
linear wave mode. It is easily shown that this result is invariant to
rotation of the wave vector $\V{k}$, along with its corresponding
linear eigenfunction, about the direction of the mean magnetic field.
The
\emph{normalized fluctuating magnetic helicity density} is defined by
\begin{equation}
\sigma_m(\V{k}) \equiv k H'_m(\V{k}) / | \V{B}(\V{k})|^2,
\label{eq:sigm1}
\end{equation}
where $k=|\V{k}|$. This normalized measure has values within the range
$[-1, +1]$, where negative values denote left-handed helicity and
positive values denote right-handed helicity.

We numerically calculate $\sigma_m(\V{k})$ over the
$k_\perp$--$k_\parallel$ plane for the eigenmodes of the linear
Vlasov-Maxwell dispersion relation \citep{Stix:1992} for a proton and
electron plasma with an isotropic Maxwellian equilibrium distribution
function for each species and no drift velocities \citep[see][for a
description of the code]{Howes:2006}.  The dispersion relation depends
on five parameters $\omega = \omega_{VM}(k_\perp \rho_i,k_\parallel
\rho_i, \beta_i, T_{i}/T_{e}, v_{th_i}/c)$, for ion Larmor radius
$\rho_i$, ion plasma beta $\beta_i$, ion to electron temperature ratio
$T_{i}/T_{e}$, and ion thermal velocity to the speed of light
$v_{th_i}/c$.

\begin{figure}
\resizebox{3.1in}{!}{\includegraphics*[0.3in,3.9in][8.0in,9.6in]{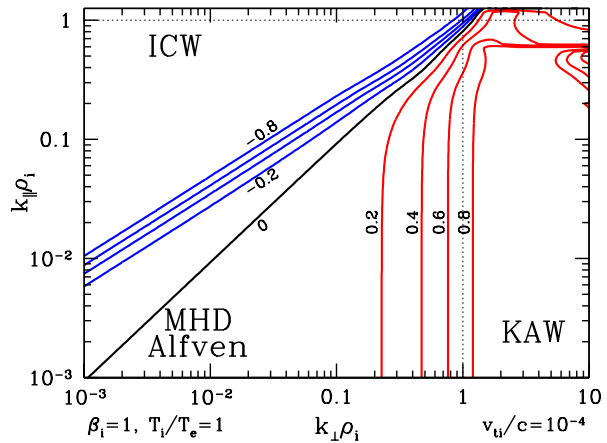}}
\caption{Normalized fluctuating magnetic helicity density
  $\sigma_m(\V{k})$ [eq. \ref{eq:sigm1}] for linear \Alfven waves over
  the $k_\perp$--$k_\parallel$ plane.  The MHD \Alfven wave (MHD
  Alfven), ion cyclotron wave (ICW), and kinetic \Alfven wave (KAW)
  regimes are identified. Plasma parameters are representative of the
  near-Earth solar wind.}
\label{fig:mhel}
\end{figure}

We specify plasma parameters characteristic of the solar wind at 1~AU:
$\beta_i=1$, $T_i/T_e=1$, and $v_{th_i}/c=10^{-4}$. \figref{fig:mhel}
is a contour plot of $\sigma_m(\V{k})$ obtained by solving for the
\Alfven wave root over the $k_\perp$--$k_\parallel$ plane, then using
the complex eigenfunctions to determine $\sigma_m(\V{k})$. The MHD
regime corresponds to the lower left corner of the plot, $k_\parallel
\rho_i \ll1$ and $k_\perp \rho_i \ll 1$; here, the \Alfven wave with
$k_\perp \sim k_\parallel$ is linearly polarized with $\sigma_m \simeq
0$. As one moves up vertically on the plot to the regime $k_\parallel
\gg k_\perp$, the solution becomes left-handed with values of
$\sigma_m \rightarrow -1$. In this regime of nearly parallel wave
vectors, the solution represents \Alfven waves in the limit
$k_\parallel \rho_i \ll \sqrt{\beta_i}$ and ion cyclotron waves in the
limit $k_\parallel \rho_i \gtrsim \sqrt{\beta_i}$. The linear wave
mode becomes strongly damped via the ion cyclotron resonance at a
value of $k_\parallel \rho_i \gtrsim \sqrt{\beta_i}$
\citep{Gary:2004}. This is precisely the behavior supporting the
cyclotron damping interpretation of the measured magnetic helicity in
the solar wind.

But the \Alfven wave solution does not always produce left-handed
magnetic helicity. If one moves instead from the MHD regime
horizontally to the right, the solution becomes right-handed with
$\sigma_m \rightarrow +1$ as $k_\perp \rho_i \rightarrow 1$,
a behavior previously found by \cite{Gary:1986}. In
this regime of nearly perpendicular wave vectors with $k_\perp \gg
k_\parallel$, the solution represents \Alfven waves in the limit
$k_\perp \rho_i \ll 1$ and kinetic \Alfven waves in the limit $k_\perp
\rho_i \gtrsim 1$. Thus, if the dissipation range is comprised of
kinetic \Alfven waves, as suggested by theories for critically
balanced, low-frequency plasma turbulence
\citep{Schekochihin:2009,Howes:2008b}, one would expect to observe a
positive normalized fluctuating magnetic helicity signature in that
regime.

\section{Reduced Fluctuating Magnetic Helicity}
\label{sec:red}
Unfortunately, due to the limitations of single-point satellite
measurements, \eqsref{eq:hm}{eq:sigm1} cannot be used directly to
calculate the fluctuating magnetic helicity from observations;
approximations must be introduced to define a related measurable
quantity. In this section, we calculate the reduced fluctuating
magnetic helicity density, as defined by \citet{Matthaeus:1982a} and
used by subsequent authors, for the magnetic field defined by
\eqref{eq:b}, but without assuming the  Taylor hypothesis.


 The two-point, two-time magnetic field correlation function is
\begin{equation}
R_{ij}(\V{r},t)=\left\langle \delta B_i(\V{x},\tau) \delta
B_j(\V{x}+\V{r},\tau+t)
\right\rangle, 
\end{equation}
where the angle brackets specify an ensemble average, defined here by
$\left\langle a(\V{r},t)\right\rangle =L^{-3}\int d^3\V{x}
a(\V{x},\V{r},t)$. We find
\begin{equation}
R_{ij}(\V{r},t)  =  \sum_{\V{k}} B_i^*(\V{k}) B_j(\V{k}) e^{i(\V{k}\cdot \V{r} - \omega  t)}
\end{equation} 
where the reality conditions ensure that this quantity is real.

We choose to sample this correlation function at a moving probe with
position given by $\V{r}=-\V{v}t$; this corresponds to satellite
measurements of the solar wind, where the probe is stationary and the
solar wind is streaming past the probe at velocity $\V{v}$.  Thus, we
may determine the reduced magnetic field correlation function,
$R^r_{ij}(t)=\left. R_{ij}(\V{r},t)\right|_{\V{r}=-\V{v}t}$, obtaining
the form
\begin{equation}
R^r_{ij}(t)=\sum_{\V{k}} B_i^*(\V{k}) B_j(\V{k}) e^{-i(\V{k}\cdot \V{v} + \omega)  t}.
\end{equation}
The reduced frequency spectrum, defined by $
S^r_{ij}(\omega')=(1/2 \pi)\int dt' R^r_{ij}(t') e^{i \omega'
t'}$, is then given by
\begin{equation}
S^r_{ij}(\omega')= \sum_{\V{k}} B_i^*(\V{k}) B_j(\V{k}) \delta[\omega'- (\V{k}\cdot \V{v}+\omega)].
\label{eq:sr}
\end{equation}
This demonstrates that the frequency $\omega'$ of the fluctuations
sampled by the moving probe is the Doppler shifted frequency
$\omega'=\V{k}\cdot \V{v} + \omega$. Note that adopting the Taylor
hypothesis \citep{Taylor:1938}, as often done in studies of solar wind
turbulence, corresponds to dropping $\omega$ in \eqref{eq:sr}.

The \emph{reduced fluctuating magnetic helicity density} is defined
by
\begin{equation}
H_m^{'r}(k_1)= 2 \mbox{Im}[S^r_{23}(k_1)]/k_1.
\label{eq:mhr}
\end{equation}
where the effective wavenumber is calculated from the measured
frequency using $ k_1 = \omega'/v$, assuming the Taylor hypothesis is
satisfied \citep{Matthaeus:1982a,Matthaeus:1982b}, and we have chosen
an orthonormal basis with direction 1 along the direction of sampling
$\hat{\V{v}}=\V{v}/|\V{v}|$ and directions 2 and 3 in the plane
perpendicular to $\hat{\V{v}}$.  The \emph{normalized reduced
  fluctuating magnetic helicity density } is given by
$\sigma^r_m(k_1)=k_1 H^{'r}_m(k_1)/|\V{B}(k_1)|^2$, where
$|\V{B}(k_1)|^2$ is the trace power.

The relation between the reduced fluctuating magnetic helicity density
$H_m^{'r}(k_1)$ and the fluctuating magnetic helicity density
$H'_m(\V{k})$ can be seen by writing the spectrum in terms of the
Doppler-shifted frequency $\omega'$ instead of $k_1$, $
H_m^{'r}(\omega') \equiv 2 \mbox{Im}[S_{23}(\omega')]/(\omega'/v)$.
Using \eqref{eq:sr} and $2\mbox{Im}[a^*b]=i(ab^* - a^*b)$, the reduced
fluctuating magnetic helicity density can be written as
\begin{eqnarray}
H_m^{'r}(\omega')&= &\sum_{\V{k}} \left(\frac{i
[B_2(\V{k})B^*_3(\V{k})- B^*_2(\V{k})B_3(\V{k})]}{\omega'/v} \right) \nonumber \\
&\times&\delta[\omega'- (\V{k}\cdot \V{v}+\omega)]
\label{eq:hmrom}
\end{eqnarray}

Equation (\ref{eq:hmrom}), the experimentally accessible quantity, is
in terms of the magnetic field measurements in a frame defined by the
solar wind velocity $\V{v}$.  To write this in terms of the
theoretically calculable $H'_m(\V{k})$ (eq. \ref{eq:hm}), we express
the magnetic field components $B_2$ and $B_3$ in the $x, y, z$
coordinate system. To do so, define the probe velocity in spherical
coordinates about the direction of the mean magnetic field: $\V{v} = v
\sin \theta \cos \phi \xhat + v \sin \theta \sin \phi \yhat + v \cos
\theta \zhat$.  The orthonormal basis specified with respect to
$\hat{\V{v}}$ can be written as
\begin{equation}
\begin{array}{ccl}
\hat{\V{e}}_1 & =&  \hat{\V{v}}  =   \sin \theta \cos \phi  \xhat 
+ \sin \theta \sin \phi \yhat + \cos \theta \zhat  \\
\hat{\V{e}}_2 & =&  \zhat \times \hat{\V{v}}  /| \zhat \times\hat{\V{v}} | =  
 -\sin \phi \xhat + \cos \phi \yhat \\
\hat{\V{e}}_3 & =& \hat{\V{e}}_1 \times \hat{\V{e}}_2  =  
-\cos \theta  \cos \phi \xhat - \cos \theta  \sin  \phi \yhat + \sin \theta  \zhat ,
\end{array}
\label{eq:basis2}
\end{equation}

Finally, we exploit the fact that the solutions of the Vlasov-Maxwell
dispersion relation depend only on the perpendicular and parallel
components of the wave vector $k_\perp$ and $k_\parallel$ with respect
to the mean magnetic field, and not on the angle about the field; thus
the eigenfunction for a wave vector $\V{k}= k_\perp \xhat +
k_\parallel \zhat$ can be rotated by an angle $\alpha$ about the mean
magnetic field to yield the solution for any wave vector $\V{k}'
=k_\perp \cos \alpha\xhat + k_\perp \sin \alpha \yhat + k_\parallel
\zhat$.
Using the above, the reduced fluctuating magnetic helicity density
$H_m^r(\omega')$ in \eqref{eq:hmrom} becomes
\begin{eqnarray}
H_m^{'r}(\omega')&=& \sum_{\V{k}} H'_m(\V{k})\frac{k_\perp \sin \theta  \cos \alpha +  k_\parallel\cos \theta }
{k_\perp \sin \theta  \cos \alpha + k_\parallel \cos \theta + \omega/v}
\nonumber \\ 
&\times & \delta[\omega'- (\V{k}'\cdot \V{v}+\omega)],
\label{eq:hmr}
\end{eqnarray}
where we have specified the azimuthal angle of the probe velocity
$\phi=0$ without loss of generality. It is clear from this equation
that all possible wave vectors $\V{k'}$ that give the same Doppler
shifted frequency $\omega'$ will contribute to the sum for the reduced
fluctuating magnetic helicity density at the frequency $\omega'$.

\section{Discussion}
\label{sec:discuss}

Predicting the values of $H_m^{'r}(\omega')$ for solar wind turbulence
based on equation (\ref{eq:hmr}) requires understanding three issues:
the scaling of the magnetic fluctuation spectrum with wavenumber, the
imbalance of \Alfven wave energy fluxes in opposite directions along
the mean magnetic field, and the variation of the angle $\theta$
between the solar wind velocity $\V{v}$ and the mean magnetic field.

The 1-D magnetic energy spectrum in the solar wind typically scales as
$k_1^{-5/3}$ in the inertial range and $k_1^p$ in the dissipation
range, where $-2\le p \le -4$ \citep{Smith:2006} and the effective
wavenumber is $k_1=\omega'/v$. It is clear from \eqref{eq:hmr} that,
when the plasma frame frequency $\omega$ is negligible, the
Doppler-shifted observed frequency always results in an effective
wavenumber $k_1 \le k$, with equality occurring only when the velocity
$\V{v}$ is aligned with the wave vector $\V{k}$. We assume that, for
homogeneous turbulence at the dissipation range scales, turbulent
energy at fixed $k_\perp$ and $k_\parallel$ is uniformly spread over
wave vectors with all possible angles $\alpha$ about the mean magnetic
field.  Because the fluctuation amplitude deceases for larger
effective wavenumbers, the contribution to $H_m^{'r}(\omega')$ is
maximum at angle $\alpha=0$; for angles $\alpha$ yielding a Doppler shift to
lower effective wavenumbers $k_1<(k_\perp^2+k_\parallel^2)^{1/2}$, the
higher amplitude fluctuations at those lower wavenumbers will
contribute more strongly to $H_m^{'r}(\omega')$.  An accurate
calculation of the magnetic helicity signature based on \eqref{eq:hmr}
must take into account the scaling of the magnetic energy spectrum.

\begin{figure}
\resizebox{3.1in}{!}{\includegraphics*[0.3in,2.in][8.0in,5.1in]{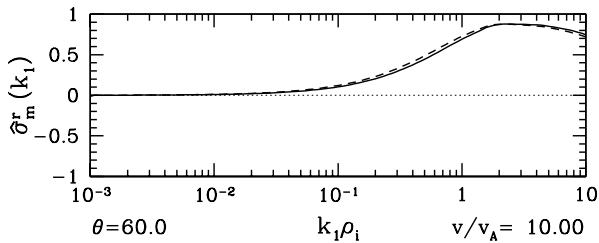}}
\caption{\label{fig:mhelr} Normalized reduced fluctuating 
magnetic helicity $\hat{\sigma}^r_m(k_1)$ vs.~effective wavenumber
$k_1$ due to a turbulent spectrum of kinetic \Alfven waves with
$\theta=60^\circ$. The solid line corresponds to the model 1-D energy
spectrum while the dashed line corresponds to a $k^{-1}$ spectrum.}
\end{figure}

To compare to $\sigma^r_m(k_1)$ derived from observations (for
example, see Figure~1 of \citet{Leamon:1998a}), we construct
the normalized quantity 
\begin{equation}
\hat{\sigma}^r_m(k_1)= \frac{\sum_{\V{k}} H'_m(\V{k})\frac{\V{k}'\cdot \V{v} }
{\V{k}'\cdot \V{v}+\omega} \delta[\omega'- (\V{k}'\cdot \V{v}+\omega)]}
{ \sum_{\V{k}} [|\V{B}(\V{k})|^2/k ]\delta[\omega'- (\V{k}'\cdot \V{v}+\omega)]}.
\label{eq:numsigm}
\end{equation}
In evaluating \eqref{eq:numsigm}, we assume a model 1-D energy
spectrum\footnote{On $150 \times 150$ logarithmic gridpoints over
$k_\perp \rho_i, k_\parallel \rho_i \in [10^{-3},10^2]$, the model
weights $B^2$ as a function of $k=({k_\perp^2+k_\parallel^2})^{1/2}$
using $B^2(k) = B_0^2 \{[ (k
\rho_i)^{-1/3} + (k \rho_i)^{4/3} ]/[1+(k \rho_i)^{2}]\}^2$.} 
that scales as $k^{-5/3}$ for $k\rho_i \ll 1$ and $ k^{-7/3}$ for
$k\rho_i
\gg 1$, consistent with theories for critically balanced turbulence
\citep{Goldreich:1995,Howes:2008b,Schekochihin:2009} and solar wind
observations \citep{Smith:2006}. In \figref{fig:mhelr}, we plot
$\hat{\sigma}^r_m(k_1)$ vs.~effective wavenumber $k_1=\omega'/v$ for a
turbulent spectrum filling the MHD \Alfven and kinetic \Alfven wave
regimes ($k_\perp>k_\parallel$ and $k_\parallel \rho_i <1$) for
$\beta_i=1$, $T_i/T_e=1$, $v_{th_i}/c=10^{-4}$, $\theta=60^\circ$, and
$v/v_A=10$. The contributions to $\hat{\sigma}^r_m(k_1)$ for all
angles $\alpha$ of each wave vector are collected in 120
logarithmically spaced bins in Doppler-shifted frequency.  The results
are rather insensitive to the scaling of the 1-D magnetic energy
spectrum over the range from $k^{-1}$ to $k^{-4}$. The solid line in
\figref{fig:mhelr} corresponds to the model spectrum assumed above, while 
the dashed line corresponds to a $k^{-1}$ energy
spectrum. \figref{fig:mhelr} demonstrates that turbulence consisting
of \Alfven and kinetic \Alfven waves produces a positive
(right-handed) magnetic helicity signature in the dissipation range at
$k_1 \rho_i \gtrsim 1$.

The analysis presented in \figref{fig:mhelr} considers only waves with
$k_\parallel >0$, so all of the waves in the summation in
\eqref{eq:hmr} are traveling in the same direction. If there were an
equal \Alfven wave energy flux in the opposite direction---a case of
balanced energy fluxes, or zero cross helicity---the net
$\hat{\sigma}^r_m(k_1)$ would be zero due to the odd symmetry of
$H'_m(\V{k})$ in $k_\parallel$. It is often observed, at scales
corresponding to the inertial range, that the energy flux in the
anti-sunward direction dominates, leading to a large normalized cross
helicity \citep{Leamon:1998b}. If this imbalance of energy fluxes
persists to the smaller scales associated with the dissipation range,
a non-zero value of $\hat{\sigma}^r_m(k_1)$ is expected. However,
theories of imbalanced MHD turbulence \citep[][ and references
therein]{Chandran:2008} predict that the turbulence is ``pinned'' to
equal values of the oppositely directed energy fluxes at the
dissipation scale. This implies that, at sufficiently high wavenumber
$k_1$, the value of $\hat{\sigma}^r_m(k_1)$ should asymptote to
zero. Thus, $\hat{\sigma}^r_m(k_1)$ in \figref{fig:mhelr} would likely
drop to zero more rapidly than shown, leaving a smaller positive net
$\hat{\sigma}^r_m(k_1)$ around $k_1 \rho_i \sim 1$, consistent with
observations \citep{Goldstein:1994,Leamon:1998a,Hamilton:2008}.  We
defer a detailed calculation of the effects of imbalance to a future
paper.

The angle $\theta$ between $\V{B}_0$ and $\V{v}$ is likely to vary
during a measurement; this angle does not typically sample its full
range $0 \le \theta \le \pi$, but has some distribution about the
Parker spiral value. Calculations of $\hat{\sigma}^r_m(k_1)$ over $0
\le \theta \le \pi$ yield results that are qualitatively similar to
\figref{fig:mhelr}, so this averaging will not significantly change
our results.

Taken together, we have demonstrated that a solar wind dissipation
range comprised of kinetic \Alfven waves produces a magnetic helicity
signature consistent with observations, as presented in
\figref{fig:mhelr}.  The underlying assumption of the cyclotron
damping interpretation of magnetic helicity measurements, an
interpretation that dominates the solar wind literature
\citep{Goldstein:1994,Leamon:1998b,Leamon:1998a,Hamilton:2008}, is the
slab model, $\V{k}=k_\parallel \zhat$ and $k_\perp=0$, i.e., purely
parallel wave vectors.  As shown in \figref{fig:mhel}, only in the
limit $k_\parallel \gg k_\perp$ does the \Alfven wave root generate a
left-handed helicity $\sigma_m \rightarrow -1$ as $k_\parallel \rho_i
\rightarrow \sqrt{\beta_i}$; in the same limit, the fast/whistler root
generates a right-handed helicity $\sigma_m \rightarrow +1$ in a
quantitatively similar manner (see Figure~9 of
\cite{Gary:1986}). Strong ion cyclotron damping of the Alfv\'en/ion
cyclotron waves as $k_\parallel \rho_i
\rightarrow 1$ \citep{Gary:2004} would leave a remaining spectrum of
right-handed fast/whistler waves, as proposed by cyclotron damping
interpretation.  However, only if the majority of the turbulent
fluctuations have $k_\parallel \gtrsim k_\perp$ is the slab limit
applicable, and only if significant energy resides in slab-like
fluctuations are the conclusions drawn about the importance of
cyclotron damping valid.
There is, on the other hand, strong theoretical and empirical support
for the hypothesis that the majority of the energy in solar wind
turbulence has $k_\perp \gg k_\parallel$ (see
\citealt{Howes:2008b} and references therein).  In this case, there is
a transition to kinetic \Alfven wave fluctuations at the scale of the
ion Larmor radius.  This Letter demonstrates that a dissipation range
comprised of kinetic \Alfven waves produces a reduced fluctuating
magnetic helicity signature consistent with observations.  


%

\acknowledgments
  G.~G.~H. thanks Ben Chandran for useful discussions.  G.~G.~H. was
  supported by the DOE Center for Multiscale Plasma Dynamics, Fusion
  Science Center Cooperative Agreement ER54785.  E.~Q. and
  G.~G.~H. were supported in part by the David and Lucille Packard
  Foundation.  E.~Q. was also supported in part by NSF-DOE Grant
  PHY-0812811 and NSF ATM-0752503.

\end{document}